\documentclass[12pt]{article}

\newcommand{\sth}{\sigma^3}
\newcommand{\stw}{\sigma^2}

\newcommand{\Ngade}{{\cal N}_{\gamma\dot{\delta}}}
\newcommand{\rconclamdondmu}{r_{c_1}...r_{c_\lambda}(r_{d_1}...r_{d_\mu})^*}
\newcommand{\Malbegade}{{\cal M}^{\alpha\dot{\beta}}_{\gamma\dot{\delta}}}
\newcommand{\Nalbegade}{{\cal N}^{\alpha\dot{\beta}}_{\gamma\dot{\delta}}}
\newcommand{\vaonajbonbk}{v^{a_1}...v^{a_j}(v^{b_1}...v^{b_k})^*}
\newcommand{\beon}{\mbox{\boldmath$e$}_1}
\newcommand{\betw}{\mbox{\boldmath$e$}_2}
\newcommand{\naonajbonbk}{n^{a_1}...n^{a_j}(n^{b_1}...n^{b_k})^*}
\newcommand{\twi}{\frac{2}{ i}}
\newcommand{\ddh}{\frac{{\rm d}}{{\rm d}h}}
\newcommand{\ddhstar}{\frac{{\rm d}}{{\rm d}h^*}}
\newcommand{\varphinp}{\varphi^{(n,p)}}
\newcommand{\vaonajvjbonbkvk}{\frac{v^{a_1}...v^{a_j} }{ \mbox{\rm v}^j}\frac{(v^{b_1}...v^{b_k})^* }{ \mbox{\rm v}^{*k}}}
\newcommand{\Ialbegade}{{\cal I}^{\alpha\dot{\beta}}_{\gamma\dot{\delta}}}
\newcommand{\hzarcz}{h(z) \propto z$ or $\arcsin z}
\newcommand{\harcz}{h(z) \propto \arcsin z}
\newcommand{\tilNgade}{\tilde{\cal N}_{\gamma\dot{\delta}}}
\newcommand{\tilIalbegade}{\tilde{\cal I}^{\alpha\dot{\beta}}_{\gamma\dot{\delta}}}
\newcommand{\hz}{h(z) \propto z}
\newcommand{\bs}{\mbox{\boldmath$s$}}
\newcommand{\bq}{\mbox{\boldmath$q$}}
\newcommand{\tilIalga}{\tilde{\cal I}^{\alpha}_{\gamma}}
\newcommand{\tilIdotbedotde}{\tilde{\cal I}^{\dot{\beta}}_{\dot{\delta}}}

\newcommand{\tilMalga}{\tilde{\cal M}^{\alpha}_{\gamma}}
\newcommand{\tilJal}{\tilde{\cal J}^{\alpha}}
\newcommand{\tilJj}{\tilde{\cal J}_j}
\newcommand{\bsv}{\mbox{\scriptsize\boldmath$v$}}
\newcommand{\bsr}{\mbox{\scriptsize\boldmath$r$}}

\newcommand{\pOmsthE}{\,^+\!\Omega_{\sigma^3{\rm E}}}

\newcommand{\mOmsthE}{\,^-\!\Omega_{\sigma^3{\rm E}}}

\newcommand{\pmOmsth}{\,^{\pm}\!\Omega_{\sigma^3}}
\newcommand{\pmOmsthE}{\,^{\pm}\!\Omega_{\sigma^3{\rm E}}}

\newcommand{\pmbvstw}{\,^{\pm}\!\mbox{\boldmath$v$}_{\sigma^2}}
\newcommand{\pmbvstwE}{\,^{\pm}\!\mbox{\boldmath$v$}_{\sigma^2{\rm E}}}
\newcommand{\pmbvE}{\,^{\pm}\!\mbox{\boldmath$v$}_{\rm E}}
\newcommand{\pbnE}{\,^{+}\!\mbox{\boldmath$n$}_{\rm E}}
\newcommand{\pbrE}{\,^{+}\!\mbox{\boldmath$r$}_{\rm E}}
\newcommand{\pmRstw}{\,^{\pm}\!R_{\sigma^2}}
\newcommand{\pmrstw}{\,^{\pm}\!r_{\sigma^2}}
\newcommand{\pbvstw}{\,^{+}\!\mbox{\boldmath$v$}_{\sigma^2}}
\newcommand{\pbvstwE}{\,^{+}\!\mbox{\boldmath$v$}_{\sigma^2{\rm E}}}
\newcommand{\pbtaustw}{\,^{+}\!\mbox{\boldmath$\tau$}_{\sigma^2}}

\newcommand{\pRstw}{\,^{+}\!R_{\sigma^2}}
\newcommand{\mbvstw}{\,^{-}\!\mbox{\boldmath$v$}_{\sigma^2}}
\newcommand{\mbvstwE}{\,^{-}\!\mbox{\boldmath$v$}_{\sigma^2{\rm E}}}
\newcommand{\bvstw}{\mbox{\boldmath$v$}_{\sigma^2}}
\newcommand{\mRstw}{\,^{-}\!R_{\sigma^2}}
\newcommand{\Rstw}{R_{\sigma^2}}
\newcommand{\taustw}{\tau_{\sigma^2}}
\newcommand{\Omsth}{\Omega_{\sigma^3}}
\newcommand{\OmT}{\Omega^{\rm T}}
\newcommand{\RT}{R^{\rm T}}

\newcommand{\pR}{\,^{+}\!R}

\newcommand{\sumstw}{\sum_{\sigma^2}}

\renewcommand{\d}{{\rm d}}
\renewcommand{\Re}{{\mbox{Re\,}}}
\renewcommand{\Im}{{\mbox{Im\,}}}
\newcommand{\D}{{\cal D}}
\newcommand{\N}{{\cal N}}
\newcommand{\F}{{\cal F}}

\newcommand{\pmbv}{\,^{\pm}\!\mbox{\boldmath$v$}}
\newcommand{\pbv}{\,^{+}\!\mbox{\boldmath$v$}}
\newcommand{\mbv}{\,^{-}\!\mbox{\boldmath$v$}}
\newcommand{\bv}{\mbox{\boldmath$v$}}
\newcommand{\rv}{{\rm v}}
\newcommand{\bu}{\mbox{\boldmath$u$}}
\newcommand{\bw}{\mbox{\boldmath$w$}}
\newcommand{\bn}{\mbox{\boldmath$n$}}
\newcommand{\bsn}{\mbox{\scriptsize\boldmath$n$}}
\newcommand{\bl}{\mbox{\boldmath$l$}}

\newcommand{\br}{\mbox{\boldmath$r$}}
\newcommand{\bvarphi}{\mbox{\boldmath$\varphi$}}
\newcommand{\bpsi}{\mbox{\boldmath$\psi$}}
\newcommand{\bphi}{\mbox{\boldmath$\phi$}}

\newcommand{\pv}{\,^{+}\!v}
\newcommand{\prvE}{\,^{+}\!{\rm v}_{\rm E}}
\newcommand{\pr}{\,^{+}\!r}

\newcommand{\mrvE}{\,^{-}\!{\rm v}_{\rm E}}
\newcommand{\gammaE}{\gamma_{\rm E}}
\newcommand{\pmrvE}{\,^{\pm}\!{\rm v}_{\rm E}}

\newcommand{\pmvstw}{\,^{\pm}\!v_{\sigma^2}}

\newcommand{\vstw}{v_{\sigma^2}}
\newcommand{\rvstw}{{\rm v}_{\sigma^2}}

\newcommand{\mR}{\,^{-}\!R}

\newcommand{\dfun}{$\delta$-function }
\newcommand{\dfuns}{$\delta$-functions }
\newcommand{\desi}{\delta^6}
\newcommand{\deth}{\delta^3}
\newcommand{\detw}{\delta^2}
\newcommand{\de}{\delta}

\newcommand{\sh}{{\rm sh}}
\newcommand{\ch}{{\rm ch}}
\newcommand{\arsh}{{\rm arsh}}

\begin{document}

\title{Defining integrals over connections in the discretized gravitational
functional integral
}
\author{V.M. Khatsymovsky \\
 {\em Budker Institute of Nuclear Physics} \\ {\em
 Novosibirsk,
 630090,
 Russia}
\\ {\em E-mail address: khatsym@inp.nsk.su}}
\date{}
\maketitle

\begin{abstract}
Integration over connection type variables in the path integral for the discrete form of the first order formulation of general relativity theory is studied. The result (a generalized function of the rest of variables of the type of tetrad or elementary areas) can be defined through its moments, i. e. integrals of it with the area tensor monomials. In our previous paper these moments have been defined by deforming integration contours in the complex plane as if we had passed to an Euclidean-like region. In the present paper we define and evaluate the moments in the genuine Minkowsky region. The distribution of interest resulting from these moments in this non-positively defined region contains the divergences. We prove that the latter contribute only to the singular (\dfun like) part of this distribution with support in the non-physical region of the complex plane of area tensors while in the physical region this distribution (usual function) confirms that defined in our previous paper which decays exponentially at large areas.

Besides that, we evaluate the basic integrals over which the integral over connections in the general path integral can be expanded.

\end{abstract}


\section{Introduction}

In the previous paper\cite{I} we have considered integration over the discrete connection type variable in the path integral for the discrete version of the first order formulation of Einstein gravity. Definition of the Euclidean version of the path integral requires special treatment because of unboundedness of the action\cite{Ham}. If we are left in the Minkowsky region, we have to carefully define integrals over the exponentially growing on the Lorentz boosts Haar measure on the discrete SO(3,1) connections. These can be given sense of the generalized functions.

The notations and motivating features of the approach are discussed in Ref. \cite{I}. Originally in the continuum formulation we would have the first order general relativity action known as the Holst action\cite{Holst} parameterized by the Barbero-Immirzi parameter\cite{Barb,Imm}. Usual perturbative definition of the functional (continual) integral does not work because of the nonrenormalizability of the theory of gravity. So we consider a minisuperspace system with finite number of the degrees of freedom. The path integral becomes usual multiple integral and can be directly evaluated. In the case of gravity we have natural minisuperspace formulation called Regge calculus or simplicial gravity\cite{Regge,Cheeger}. This formulation deals with the piecewise flat spacetime (collection of the flat four-dimensional tetrahedra or 4-simplices) whose geometry is completely described by the edge lengths. It has an important property of being able to approximate in some sense arbitrarily accurately any given arbitrary curved spacetime when triangulation (edge) length dynamical variables take on their values in certain region (are sufficiently small). So any quantum amplitude is defined, among others, also by paths passing arbitrarily closely to any given arbitrary spacetime. That is, we can expect that using the Regge minisuperspace on quantum level we do not lose essential degrees of freedom connected with non-piecewise-flat spacetimes. Methodologically, the simplicial gravity resembles field theory on the {\it dynamical} lattice whose spacings (edge lengths) are dynamical variables themselves. It would be of significance for consistency and finiteness of the theory if, roughly speaking, the quantum probability distribution for the simplicial edge lengths were concentrated around certain finite nonzero values of the order of Plank scale. Then we would have something like usual lattice but with spacings fixed {\it dynamically}.

Invoking the notion of discrete tetrad and connection first considered in Ref. \cite{Fro} we have suggested in Ref. \cite{Kha} representation of the Einstein action $\frac{1}{2} \int{R\sqrt{-g}{\rm d}^4x}$ on the piecewise flat manifold in terms of selfdual and antiselfdual parts of area tensors and of finite rotation SO(4) (SO(3,1) in the Minkowsky case) matrices.

Integration over these connection type variables in the path integral studied just has relevance to the probability distribution for the simplicial lengths. The selfdual and antiselfdual representations of Ref. \cite{Kha} can be generally combined into the one-parametric family of real representations for the action,
\begin{eqnarray}\label{S}                                                          
S = \frac{1}{2} \sumstw \left [ \left (1 + \frac{i}{\gamma} \right ) \sqrt{\pbvstw^2} \arcsin \frac{\pbvstw * \pRstw (\{ \Omsth \} )}{\sqrt{\pbvstw^2} } \right. \nonumber\\ \left. + \left (1 - \frac{i}{\gamma} \right ) \sqrt{\mbvstw^2} \arcsin \frac{\mbvstw * \mRstw (\{ \Omsth \} )}{\sqrt{\mbvstw^2} } \right ]
\end{eqnarray}

\noindent where $\gamma$ is the discrete analog of the Barbero-Immirzi parameter denoted by the same letter. Here $\pbvstw$ and $\mbvstw = (\pbvstw )^*$ are the 3-vectors parameterizing (anti-)selfdual parts $\pmvstw^{ab}$ of area tensor $\vstw^{ab}$ of the triangle $\stw$. If $v^{ab} = \frac{1}{2}\epsilon^{ab}_{~~cd}l^c_1l^d_2$ (bivector) for some two 4-vectors $l^c_1$, $l^d_2$ which span the triangle, then $2 \pmbv = \pm i \bl_1 \times \bl_2 - \bl_1 l^0_2 + \bl_2 l^0_1$, and $\pmbv^2$ is $(-1)$ times the square of the (real for the spacelike triangle) area. The $\Omsth$ is SO(3,1) matrix on the tetrahedron $\sth$ which we call simply connection. The $\Rstw$ is curvature matrix on the triangle $\stw$ (holonomy of $\Omega$s). For a 3-vector $\bv$ and a $3\times 3$ matrix $R$ we have denoted $\bv * R \equiv \frac{1}{2}v^a R^{bc} \epsilon_{abc}$, and for $\pRstw$ and $\mRstw = (\pRstw )^*$, the (multiplicative) (anti-)selfdual parts of $\Rstw$, as it is seen, we have used adjoint, SO(3) representation (to be precise, SO(3,C) matrix).

Now we write out the functional integral mentioned, $\int \exp (iS) D q$, $q$ are field variables (some factors of the type of Jacobians could also be present). Functional integral approach in Regge calculus was earlier developed, see, e. g., Refs. \cite{Fro,HamWil1,HamWil2}. Suppose we have performed integration over connections and are interested in the dependence of the intermediate result on area tensors. Of course, the different (components of) bivectors are not independent, but nothing prevents us from studying analytical properties in the extended region of varying area tensors as if these were independent variables more general than bivectors. Namely, we consider integral
\begin{equation}\label{N-omega}                                                    
\N = \int \exp (iS ) \prod_{\sth} \D \Omsth.
\end{equation}

The integrand in Eq. (\ref{N-omega}) effectively depends on $\Omsth$s through their products $\Rstw$s. The latter on a certain set of the triangles $\F$ are functions of others, $\{ \Rstw : \stw \not \in \F \}$, due to the Bianchi identities\cite{Regge}. The overall number of the connections $\Omsth$ is larger than the number of independent curvatures $\{ \Rstw : \stw \not \in \F \}$ by a number of some connections taken as purely gauge ones. Fixing the latter, $\Omsth = 1$, means extension of any 4-simplex local frame to the global one across these $\sth$s while other $\sth$s are viewed in two copies separated by the cuts along them (or, equally to say, by the single multi-branch cut). Any such cut (or, equally to say, a branch of the multi-branch cut) is made along a chain of the tetrahedrons $\sth_k$, $k = 1, 2, \dots$ passing successively through the triangles $\sth_{k - 1} \cap \sth_k = \stw_k$ so that
\begin{eqnarray}\label{R-omega-int}                                              
& & R_{\stw_k} = \OmT_{\sth_{k - 1}} \Omega_{\sth_k}, ~~~ k = 2, 3 \dots , \\
& & R_{\stw_1} = \Omega_{\sth_1}
\end{eqnarray}

\noindent and
\begin{equation}\label{omega-R}                                                    
\Omega_{\sth_n} = R_{\stw_1} R_{\stw_2} \dots R_{\stw_n}, ~~~ n = 1, 2, \dots
\end{equation}

\noindent (may be, up to possible change $R_{\stw} \to \RT_{\stw}$, $\Omega_{\sth} \to \OmT_{\sth}$ for some $\stw$s, $\sth$s). Then $\D \Omega_{\sth_n} = \D R_{\stw_n}$ due to the invariance of the Haar measure. Therefore, up to integrations over the connections taken as purely gauge ones, the integration element $\prod_{\sth} \D \Omega_{\sth}$ is reduced to that one over independent curvatures, $\prod_{\stw \not \in \F} \D R_{\stw}$.

It quite may be that a triangle $\stw$ cannot be written as $\sth_{k - 1} \cap \sth_k$ for a chain of tetrahedrons $\sth_k$, $k = 1, 2, \dots$ defining a cut. By definition, every tetrahedron, as well as any one containing this triangle, $\sth \supset \stw$, either belongs to some cut or $\Omsth$ is gauge connection, $\Omsth = 1$. In the former case the $\Omsth$ is by Eq. (\ref{omega-R}) a function of the curvatures on the triangles {\it other} than the given $\stw$. The resulting curvature $R_{\stw}$ is a function of the curvatures on such triangles too. Therefore the triangle $\stw$ belongs to the set $\F$ on which the curvatures are functions of the independent ones, $R_{\stw} (\{ \Rstw : \stw \not \in \F \})$. Redenote area tensors of the triangles of $\F$ as $\vstw \to \taustw$. So we get (up to integrations over the purely gauge $\Omega_{\sth}$s)
\begin{eqnarray}\label{N-R}                                                        
& & \N = \int \exp \frac{i}{2} \left \{ \left (1 + \frac{i}{\gamma} \right ) \left [ \sum_{\stw \not \in \F} \sqrt{\pbvstw^2} \arcsin \frac{\pbvstw * \pRstw }{\sqrt{\pbvstw^2} } \right. \right. \nonumber\\ & & \left. + \sum_{\stw \in \F} \sqrt{\pbtaustw^2} \arcsin \frac{\pbtaustw * \pRstw (\{ \Rstw : \stw \not\in {\cal F}\} ) }{\sqrt{\pbtaustw^2} } \right ] \nonumber \\ & & \left. \phantom{\frac{\sqrt{1}}{\sqrt{1}}} + \mbox{complex conjugate} \right \} \prod_{\stw \not \in \F} \D \Rstw.
\end{eqnarray}

\noindent In the consistent path integral approach in the theory with independent area tensors the $\taustw$s should be considered as fixed\cite{I}. (These are the discrete analogs of the gauge fixed (non-dynamical) variables in the canonical formalism resulting in the continuous time limit. The triangles of $\F$ can be described as those containing the lapse-shift edges being the discrete analogs of the lapse-shift vectors.)

Example of the simplicial structure quite universal for approximating any smooth geometry is that constructed of the sequence of the same three-dimensional simplicial structures (leaves). The construction proceeds by adding the new links connecting the vertices of the each two neighboring leaves. Namely, each two analogous vertices in both leaves are connected; besides that, we add one of the two possible diagonals in the each quadrangle formed by two analogous edges in the both leaves (and by two else links connecting analogous ends of these edges between the leaves). Thereby, in particular, the space between the two neighboring leaves is divided into the four-dimensional prisms with bases being the 3-simplices $\sth$ of these leaves. The overall complex is divided into the four-dimensional prisms. The above purely gauge connections "live" on the internal 3-simplices $\sth$ of the 4-prisms. (The 4-prism does not contain internal curvature and can be considered as flat.) The cuts can be made along the three-dimensional lateral faces of the 4-prisms, the three-dimensional prisms themselves. The triangles $\stw \in \F$ are those belonging to the two-dimensional lateral faces of the 3-prisms, i. e. to the two-dimensional prisms or bands consisting of the triangles.

Some features of the behavior of the expressions like (\ref{N-R}) can be illustrated by the following model integral,
\begin{equation}\label{model}                                                      
\int\limits^{+\infty}_{-\infty} e^{iA \arsh (\lambda \sh \psi )} \d \psi = 2 \pi \de (A) + (1 - \lambda^2) \frac{\pi A / 2}{\sh (\pi A / 2 )} F (1 + iA / 2 , 1 - iA / 2 ; 2 ; 1 - \lambda^2 ).
\end{equation}

\noindent Here $A$ is modelling value of the spacelike area, $\sqrt{- \pmbvstw^2}$; $\psi$ is modelling a Lorentz boost angle; $\lambda$ is modelling $\cos \theta$, $\theta$ being the angle between $\pmbvstw$ and the vector $\pmrstw^a = \epsilon^a_{bc} \pmRstw^{bc} / 2$ directed along the axis of the rotation $\pmRstw$. The Eq. (\ref{model}) follows after passing to the new variable $x = \arsh (\lambda \sh \psi )$ and expanding in powers of $1 - \lambda^2$. The known asymptotic of the hypergeometric function at large parameters\cite{Erdel}, $A \to \infty$, gives
\begin{eqnarray}                                                                   
& & \int\limits^{+\infty}_{-\infty} e^{iA \arsh (\lambda \sh \psi )} \d \psi = 2 \pi \de (A) + \sqrt{2 \pi} \frac{(1 - \lambda^2)^{1/4}}{\sqrt{\lambda | A | }} e^{- | A | \arcsin \lambda } \left [ 1 \right. \nonumber \\ & & \left. + O( A^{-1}) + O(e^{-2 | A | \arccos \lambda }) \right ], ~~~ 0 < \lambda < 1.
\end{eqnarray}

\noindent The exponential suppression observed at large areas is favorable for the internal consistency of the simplicial description. However, occurrence of the \dfun $\de (A)$ is not good for the consistency for this would mean either degenerate configuration or returning to the continuum description formally just corresponding to the infinitely fine triangulation $A \to 0$.

Remarkable is that the actual expression (\ref{N-R}) possesses, apart from the exponential suppression at large areas, also the $\delta$-function-like terms with support {\it shifted away} from the physical region $\Im \pmbvstw^2 = 0$ ($\vstw^{ab}$ are bivectors).

The present paper just defines and calculates the basic integrals over which the expression (\ref{N-R}) could be expanded. This proceeds, as in the paper \cite{I}, through defining integrals of the functions of interest with area tensor monomials, the so-called moments. Whereas the moments in the paper \cite{I} were defined and evaluated by passing to the Euclidean-like region, now we perform direct calculation in the genuine Minkowsky region, analyze the arising divergences and establish identity of the both approaches in the physical region.

\section{Defining the moments of the path integral distribution}

As mentioned above, we regard Eq. (\ref{N-R}) as function of arbitrary $\pbv$, $\mbv = (\pbv)^*$ which we redenote as $\bv, \bv^*$ in the main body of the paper. To be specific, we study the following integrals,
\begin{equation}\label{N-gamma-delta}                                              
\Ngade (\bv,\bv^*) \equiv \int \exp \left [\frac{i}{2}\rv h(\bn\br) + \frac{i}{2}\rv^*h(\bn\br)^*\right ] \rconclamdondmu \D R.
\end{equation}

\noindent Here $\gamma = (c_1...c_\lambda)$, $\dot{\delta} = (\dot{d}_1 ... \dot{d}_\mu)$ are multiindices; the dot on an index has the only sense that corresponding vector component enters complex conjugated. The $h(z)$ is analytical at $z=0$ odd function $h(z) = -h(-z)$. Principal value $\arcsin z$ or simply $z$ are examples of $h(z)$. Besides that, $\rv = \sqrt{\bv^2}$, $\bn = \bv / \rv$, $\bn^2 = 1$, $r_a = \epsilon_{abc}\pR^{bc}/2 = \phi_a(\sin \phi)/\phi$, $\phi = \sqrt{\bphi^2}$, $\bphi = \bvarphi - i\bpsi$,
\begin{equation}\label{DR}                                                          
\D R = \left (\frac{1}{\sqrt{1-\br^2}} - 1\right )\left (\frac{1}{\sqrt{1-\br^{*2}}} - 1\right ) \frac{\d^3\br \d^3\br^*}{(8\pi^2)^2 \br^2 \br^{*2}} \equiv \D \pR \D \mR.
\end{equation}

\noindent Here $\d^3\br \d^3\br^* \equiv 2^3 \d^3 \Re \br \d^3 \Im \br \equiv 2^3 \d^6 \br$. The monomial $\rconclamdondmu$ originates as a term in Taylor expansion of possible dependence on $R$ of the factors provided by $\Rstw$ in other triangles due to the Bianchi identities.

Generally Eq. (\ref{N-gamma-delta}) is a distribution (generalized function) and has sense being integrated with the suitable probe functions. For the latter we chose those ones for which corresponding integrals could be easily defined.
Let us issue from the integral of $\Ngade$ with powers of $\bv, \bv^*$,
\begin{equation}\label{Malbegade}                                                 
\Malbegade (l,m) = \int \Ngade (\bv, \bv^*)(\bv^2)^l (\bv^{*2})^m \vaonajbonbk \d^6 \bv,
\end{equation}

\noindent and change overall integration order: first integrate over $\d^6 \bv$, then over $\d^6 \br$. Here $\d^6 \bv \equiv \d^3 \Re \bv \d^3 \Im \bv$, etc. The $\alpha, \dot{\beta}$ are multiindices. The only sense of distinguishing between superscripts and subscripts is that the former refer to $\bv, \bv^*$, the latter refer to $\br, \br^*$. Call (\ref{Malbegade}) the moment of $\Ngade$ (specified by $\alpha, \beta, l, m$).

At $h(z) \propto z$ when calculating $\Malbegade (l, m)$ Eq. (\ref{Malbegade}) we get derivatives of the \dfun $\desi (\br )\!$ $\!\equiv\!$ $\!\deth (\Re \br)\!$ $\!\deth (\Im \br)\!$ which are then integrated over $\D R$. Finiteness is provided by analyticity of this measure at $\br, \br^* \to 0$ w.r.t. $\br, \br^*$ viewed as independent complex variables, $\D R = |c_0 + c_1 \br^2 + c_2 (\br^2)^2 + ... |^2 \d^3\br \d^3\br^*$.

In general case $h(z) \neq const \cdot z$ integral also can be defined. Again, consideration goes through intermediate appearance of \dfuns . For that we make use of special structure of the exponential in (\ref{N-gamma-delta}) and temporarily pass to components of $\bv$ which remind spherical ones, but are modified for complex case,
\begin{equation}                                                                  
\bv = \rv \bn, ~~~ \rv = u + iw, ~~~ \bn = \beon \ch \rho + i \betw \sh \rho, ~~~ \beon^2 = 1 = \betw^2, ~~~ \beon \betw = 0.
\end{equation}

\noindent The orthogonal pair $\beon, \betw$ is specified by three angles, e. g. by azimuthal $\theta_1$ and polar $\varphi_1$ angles of $\beon$ and polar angle $\varphi_2$ of $\betw$ (in the plane orthogonal to $\beon$). The integration measure in the coordinates $u, w, \rho, \theta_1, \varphi_1, \varphi_2$ is
\begin{equation}                                                                  
\d^6 \bv = (u^2 + w^2)^2\d u \d w \d^4 \bn, ~~~ \d^4 \bn = \ch \rho \sh \rho \d \rho \sin \theta_1 \d \theta_1 \d \varphi_1 \d \varphi_2.
\end{equation}

\noindent Unlike the Euclidean case, $\bn$ varies in the noncompact region. Therefore when we consider integrals over $\d^4 \bn$ at an intermediate stage below, we imply these being temporarily regularized by, e. g., requiring $|\rho | < \Lambda$ at some large but finite $\Lambda$. The $u, w$ are defined via $(u + iw)^2 = \bv^2$, i. e. the region of variation for $u + iw$ is a half of the complex plane. For example, for the standard choice of the cut for square root function $u \geq 0$. However, integration over $\d u$ in (\ref{Malbegade}) can be extended to the full real axis $(-\infty, +\infty)$. This is only possible because formal putting $u \to -u, w \to -w$ is equivalent to $\beon \to -\beon, \betw \to -\betw$ in (\ref{Malbegade}) due to the oddness of $h(z)$. Such identity of integration points leads to \dfuns of $h$,
\begin{eqnarray}\label{intvh}                                                     
 & & \int \exp \left [\frac{i}{ 2}\rv h(\bn\br) + \frac{i}{ 2}\rv^*h(\bn\br)^*\right ] (\bv^2)^l (\bv^{*2})^m \vaonajbonbk \d^6 \bv \nonumber\\ & & \hspace{-1cm} = \frac{1}{ 2} \int \naonajbonbk \d^4 \bn \int\limits^{+\infty}_{-\infty} \d u \int\limits^{+\infty}_{-\infty} \d w \, e^{i[uf(\bsn\bsr) + wg(\bsn\bsr)]} \nonumber\\ & & \cdot (u + iw)^{j+2l+2}(u - iw)^{k+2m+2} \nonumber\\
& & \hspace{-10mm} = \frac{1}{ 2} \int \naonajbonbk \d^4 \bn (2\pi)^2 \left (\twi \ddh \right )^{j+2l+2} \left (\twi \ddhstar \right )^{k+2m+2} \detw (h)
\end{eqnarray}

\noindent where
\begin{eqnarray}                                                                  
& & f(z) = \Re h(z), g(z) = -\Im h(z), 2\de (h) \de (h^*) \equiv \de (f) \de (g) \equiv \detw (h), \\                                                                           
& & \frac{\partial }{ i\partial f} + \frac{\partial }{ \partial g} \equiv \twi \ddh, ~~~
\frac{\partial }{ i\partial f} - \frac{\partial }{ \partial g} \equiv \twi \ddhstar.
\end{eqnarray}

The derivatives $\de^{(j+2l+2)}(h) \de^{(k+2m+2)}(h^*)$ can be expanded into combinations of the derivatives of $\detw (z) \equiv \de (x) \de (y)$, $z \equiv  \bn \br = x - iy$. These combinations can be found by integrating $\de^{(j+2l+2)}(h) \de^{(k+2m+2)}(h^*)$ with probe functions $\varphi (z, z^*)$. Let us choose
\begin{equation}                                                                  
\varphi (z, z^*) = \frac{\varphinp (0,0) }{ n!p!} z^n z^{*p},
\end{equation}

\noindent thus we find coefficient of $\de^{(n)}(z)\de^{(p)}(z^*)$ in $\de^{(j+2l+2)}(h) \de^{(k+2m+2)}(h^*)$,
\begin{eqnarray}\label{delta-h-to-z}                                              
\de^{(j+2l+2)}(h) \de^{(k+2m+2)}(h^*) = ... + (-1)^{j+n+k+p} \de^{(n)}(z)\de^{(p)}(z^*) \nonumber\\ \cdot \left (\ddh \right )^{j+2l+3}\left (\ddhstar \right )^{k+2m+3} \left [ \frac{z(h)^{n+1}}{ (n+1)!}\frac{z(h)^{*p+1}}{ (p+1)!} \right ]_{h,h^*=0} + ... .
\end{eqnarray}

\noindent (This term is nonzero at $(j-n)({\rm mod}2) = 0$, $(k-p)({\rm mod}2) = 0$, $j + 2l + 2 \geq n$, $k + 2m + 2 \geq p$.) Let us use formula (\ref{intvh}) read from right to left, now at $h(z) = z$,
\begin{eqnarray}\label{delta-back}                                                
& & \frac{1}{ 2} \int \naonajbonbk \d^4 \bn (2\pi)^2 \left (\twi \right )^{n + p} \de^{(n)} (\bn \br)\de^{(p)}((\bn \br)^*) \nonumber \\
& & = \int \exp \left (\frac{i}{ 2}\bv\br + \frac{i}{ 2}\bv^*\br^*\right ) \rv^{n-j} \rv^{*p-k} \vaonajbonbk \frac{\d^6 \bv }{ \rv^2 \rv^{*2}}.
\end{eqnarray}

\noindent Summation over $n$, $p$ with the coefficients found in (\ref{delta-h-to-z}) gives us (\ref{intvh}) which being integrated over $\rconclamdondmu \D R$ yields the moment
\begin{eqnarray}\label{M-I}                                                       
& & \Malbegade (l,m) = \int \rconclamdondmu \D R ~ (-1)^{j+k} \left ( \twi \ddh \right )^{j+2l+2} \left ( \twi \ddhstar \right )^{k+2m+2} \nonumber\\ & & \cdot \left \{ \frac{\d z }{ \d h} \frac{\d z^* }{ \d h^*} \int \exp \left (\frac{i}{ 2}\bv\br + \frac{i}{ 2}\bv^*\br^*\right ) \vaonajvjbonbkvk \right. \nonumber\\ & & \cdot \left. \frac{\d^6 \bv }{ \rv^2 \rv^{*2}} \left [ ... + \frac{\left (\rv z/(2i)\right )^n }{ n!} \frac{\left (\rv^*z^*/(2i)\right )^p }{ p!} + ... \right ]  \right \}_{h,h^* = 0} \nonumber \\ & & = \left ( 2i \ddh \right )^{j+2l+2} \left ( 2i \ddhstar \right )^{k+2m+2} \left [ \frac{\d z }{ \d h} \frac{\d z^* }{ \d h^*} \Ialbegade (z,z^*) \right ]_{h,h^* = 0}
\end{eqnarray}

\noindent with the "generating function"
\begin{eqnarray}\label{Ialbegade}                                                 
& & \Ialbegade (z,z^*) = \int \rconclamdondmu \D R \int \exp \left (\frac{i}{ 2}\bv\br + \frac{i}{ 2}\bv^*\br^*\right ) \vaonajvjbonbkvk \nonumber\\ & & \cdot \left [ \frac{1}{ 2} \exp \frac{\rv z }{ 2i} + \frac{(-1)^j }{ 2} \exp \frac{i\rv z }{ 2}\right ] \left [ \frac{1}{ 2} \exp \frac{\rv^*z^* }{ 2i} + \frac{(-1)^k }{ 2} \exp \frac{i\rv^*z^* }{ 2}\right ] \frac{\d^6 \bv }{ \rv^2 \rv^{*2}}.
\end{eqnarray}

\noindent We have extended summation to infinite set of the nonnegative integers $n, p$ keeping in mind that upon applying $(\d / \d h)^{j + 2l + 2} (\d / \d h^*)^{k + 2m + 2} (\cdot )_{h, h^* = 0}$ only finite number of terms at the given finite $j, k, l, m$ (pointed out after formula (\ref{delta-h-to-z})) are active.

\section{Isolating the divergences}\label{isolating}

Generally there are divergences in the integral over $\bv$ from small $\rv$ in the definition (\ref{Ialbegade}) when we remove the above mentioned regularization $|\rho | \! $ $ \! < \! $ $ \! \Lambda$ (or $[\Im (\bv /\rv)]^2 \! $ $\! < \! $ $\! (\sh \Lambda)^2$) for the noncompact angle parameter $\rho$ in the spacetime with non-positively defined metric, $\Lambda \to \infty$. Here we find that at $\hzarcz$ these divergences do not effect the physical value of $\Ngade (\bv , \bv^* )$ recovered from the moments. In particular, at $\harcz$ these divergences can be attributed to the terms in $\Ngade (\bv, \bv^*)$ having support on the discrete set of points $\rv^2 = 4 n^2 (1 + i / \gamma )^{-2}$ with integer $n > 0$. These points lay outside the physical region $\Im \rv^2 = 0$.

Namely, consider the terms $z^nC_n(z^*)$ or $C_n(z)z^{*n}$ in $\Ialbegade (z,z^*)$ with holomorphic $C_n(z)$.

\medskip

\noindent LEMMA 1. {\it At $\harcz$ the term $z^nC_n(z^*)$ or $C_n(z)z^{*n}$ at a nonnegative integer $n$ with $C_n(z)$ holomorphic at $z = 0$ in $\Ialbegade (z,z^*)$ corresponds to the term in $\Ngade (\bv, \bv^*)$ with support at the points $\rv^2 = 4\tilde{n}^2 (1 + i/ \gamma )^{-2}$, $\tilde{n} = n + 1, n - 1, ..., n({\rm mod}2) + 1$.}

\medskip

{\it Proof.} Consider the term $z^nC_n(z^*)$. The dependence of the corresponding contribution to $\Malbegade (l,m)$ on $l$ decouples as
\begin{equation}\label{z-nC-n}                                                    
\left (2i \ddh \right )^{j+2l+2} \left ( \frac{\d z }{ \d h } z^n \right )_{h=0} = \left. \left (2i \ddh \right )^{j+2l+3} \frac{z^{n+1} }{ 2i(n+1)} \right |_{h=0}.
\end{equation}

\noindent At $z = \sin \frac{h }{ 1 + i/ \gamma }$ the power $z^{n+1}$ contains harmonics $\sin \frac{\tilde{n} h }{ 1 + i/ \gamma }$ or $\cos \frac{\tilde{n} h }{ 1 + i/ \gamma }$, $\tilde{n} = n + 1, n - 1, ..., n({\rm mod}2) + 1$, for even or odd $n$, respectively. Contribution to the moment from harmonic $\sin \frac{\tilde{n} h }{ 1 + i/ \gamma }$ or $\cos \frac{\tilde{n} h }{ 1 + i/ \gamma }$ is proportional to $\frac{1}{ 2i(n+1)} \left ( \frac{2i\tilde{n} }{ 1+i/\gamma} \ddh \right )^{j+2l+3} (\sin h)$ or $\frac{1}{ 2i(n+1)} \left ( \frac{2i\tilde{n} }{ 1+i/\gamma} \ddh \right )^{j+2l+3} (\cos h)$ at $h = 0$, respectively. Nonzero contribution to the moment follows for even or odd $j$, respectively, and the dependence on $l$ is proportional to $[2\tilde{n} / (1 + i/ \gamma )]^{2l}$. On the other hand, we can combine the moments $\Malbegade (l, m)$ with different $l, m$ into the functional
\begin{equation}                                                                  
\Nalbegade (f(\rv^2) g(\rv^2)^*) = \int \Ngade (\bv, \bv^*) f(\rv^2) g(\rv^2)^* \vaonajbonbk \d^6 \bv
\end{equation}

\noindent on the polynomials $f(\rv^2)$, $g(\rv^2)^*$. Then the considered contribution to $\Nalbegade (f(\rv^2)\!$ $\! g(\rv^2)^*)$ is proportional to $f(4\tilde{n}^2 (1 + i/ \gamma )^{-2})$. This just corresponds to the term in $\Ngade (\bv, \bv^*)$ with support at $\rv^2 = 4\tilde{n}^2 (1 + i/ \gamma )^{-2}$.
\vspace{-11mm} \begin{flushright} $\spadesuit$ \end{flushright} \vspace{-1mm}

Below we find that the moments can be made well-defined by subtracting certain terms $z^nC_n(z^*)$ or $C_n(z)z^{*n}$ from $\Ialbegade (z,z^*)$. These terms correspond to the terms subtracted from $\Ngade (\bv, \bv^*)$ with support at $\rv^2 = 4\tilde{n}^2 (1 + i/ \gamma )^{-2}$, $\tilde{n} = n + 1, n - 1, ..., n({\rm mod}2) + 1$ where $n$ takes on the values of $\lambda$ and $\mu$, the lengths of the multiindices $\gamma$ and $\delta$. Thus the resulting $\tilNgade$ coincides with $\Ngade$ outside these points. It turns out that corresponding $\tilIalbegade$ factorizes into holomorphic and antiholomorphic parts. Each of these parts can be regarded as continuation of the purely real (Euclidean type) integral.

A priori $\Ngade (\bv , \bv^*)$ can be represented as linear combination of all independent tensor structures with multiindices $\gamma$, $\dot{\delta }$ composed of $v_c$, $v^*_d$, $\delta_{cd}$. It is natural that to define coefficients at these structures it is sufficient to know the results of contraction of $\Ngade$ with all these structures. This is covered by knowledge of $\Malbegade (l, m)$ at $\alpha$, $\beta$ with the lengths of these multiindices $j$, $k$ the same as the lengths $\lambda$, $\mu$, respectively. Below we imply the moments with just these lengths of $\alpha$, $\beta$.

\medskip

\noindent LEMMA 2. {\it The $\Ngade (\bv, \bv^*)$ at $\hzarcz$ (at $\harcz$ modified by subtracting certain singular terms with support at $\rv^2 = 4\tilde{\jmath}^2 (1 + i/ \gamma )^{-2}$, $\tilde{\jmath} = j + 1, j - 1, ..., j({\rm mod}2) + 1$ and $\rv^2 = 4\tilde{k}^2 (1 + i/ \gamma )^{-2}$, $\tilde{k} = k + 1, k - 1, ..., k({\rm mod}2) + 1$) possesses the moments admitting removing the regularization, $\Lambda \to \infty$, with $\Ialbegade (z, z^*)$ factorized into holomorphic and antiholomorphic parts.}

\medskip

{\it Proof.} Let us subtract from $\exp (\pm i\rv z/2)$ and from $\exp (\pm i\rv^*z^*/2)$ in square brackets in the formula (\ref{Ialbegade}) for $\Ialbegade (z,z^*)$ the first up to $\propto (\rv z)^j$ inclusive and the first up to $\propto (\rv^*z^*)^k$ inclusive terms of the Taylor expansions of these functions over $z$ and over $z^*$, respectively. The $\Ialbegade (z,z^*)$ is replaced by
\begin{eqnarray}\label{tilIalbegade}                                              
& & \tilIalbegade (z,z^*) = \int \rconclamdondmu \D R \int \exp \left (\frac{i}{ 2}\bv\br + \frac{i}{ 2}\bv^*\br^*\right ) \vaonajvjbonbkvk \nonumber\\ & & \cdot \left [ \frac{1}{ 2} \exp \frac{\rv z }{ 2i} + \frac{(-1)^j }{ 2} \exp \frac{i\rv z }{ 2} - \sum^{[j/2]}_{n=0} \frac{1 }{ (j - 2n)! } \left (\frac{\rv z }{ 2i}\right )^{j-2n}\right ] \nonumber \\ & & \cdot \left [ \frac{1}{ 2} \exp \frac{\rv^*z^* }{ 2i} + \frac{(-1)^k }{ 2} \exp \frac{i\rv^*z^* }{ 2} - \sum^{[k/2]}_{p=0} \frac{1 }{ (k - 2p)! } \left (\frac{\rv^*z^* }{ 2i}\right )^{k-2p}\right ] \frac{\d^6 \bv }{ \rv^2 \rv^{*2}}.
\end{eqnarray}

\noindent At $\harcz$ use lemma 1: this generating function defines moments of some $\tilNgade (\bv, \bv^*)$ which coincides with $\Ngade (\bv, \bv^*)$ in the region with nonphysical points (at $\harcz$) $\rv^2 = 4\tilde{\jmath}^2 (1 + i/ \gamma )^{-2}$, $\tilde{\jmath} = j + 1, j - 1, ..., j({\rm mod}2) + 1$ and $\rv^{*2} = 4\tilde{k}^2 (1 - i/ \gamma )^{-2}$ or $\rv^2 = 4\tilde{k}^2 (1 + i/ \gamma )^{-2}$, $\tilde{k} = k + 1, k - 1, ..., k({\rm mod}2) + 1$ excluded. At $\hz$ note that, e. g., the term $z^nC_n(z^*)$ in $\Ialbegade$ at $n \leq j$ does not contribute to the moments ((\ref{z-nC-n}) vanishes).

Expansion over $z, z^*$ gives nonnegative powers of $\rv^2, \rv^{*2}$,
\begin{eqnarray}                                                                  
& & \tilIalbegade (z, z^*) = \left ( \frac{z }{ 2i} \right )^{j+2} \left ( \frac{z^* }{ 2i} \right )^{k+2} \int \rconclamdondmu \D R \int \exp \left ( \frac{i}{ 2} \bv \br + \frac{i}{ 2} \bv^* \br^*\right ) \nonumber \\ & & \cdot \left [ \sum^{\infty}_{n=0} \frac{(\rv z/2i)^{2n} }{ (j+2+2n)!}\right ] \left [ \sum^{\infty}_{p=0} \frac{(\rv^*z^*/2i)^{2p} }{ (k+2+2p)!}\right ] \vaonajbonbk \d^6 \bv,
\end{eqnarray}

\noindent so that we can remove the regularization, $\Lambda \to \infty$, explicitly performing integration over the whole complex plane of $\bv$,
\begin{eqnarray}\label{factoriz}                                                  
& & \int \rconclamdondmu \D R \int \exp \left ( \frac{i}{ 2} \bv \br + \frac{i}{ 2} \bv^* \br^*\right ) (\bv^2)^n (\bv^{*2})^p \vaonajbonbk \nonumber \\ & & \cdot \d^6 \bv = \int \rconclamdondmu \D R \int \exp (i\bu\bs + i\bw\bq ) [(\bu + i \bw )^2]^n [(\bu - i \bw )^2]^p \nonumber \\
& & \cdot (u^{a_1} + iw^{a_1}) ... (u^{a_j} + iw^{a_j}) (u^{b_1} - iw^{b_1}) ... (u^{b_k} - iw^{b_k}) \d^3 \bu \d^3 \bw \nonumber \\
& & = \int \rconclamdondmu \D R \,\,\, (2\pi)^6 \left [ \left ( \frac{\partial }{ i\partial \bs} + \frac{\partial }{ \partial \bq} \right )^2 \right ]^n \left [ \left ( \frac{\partial }{ i\partial \bs} - \frac{\partial }{ \partial \bq} \right )^2 \right ]^p \nonumber \\
& & \hspace{0mm} \cdot \left ( \frac{ \partial }{ i\partial s_{a_1} } + \frac{ \partial }{ \partial q_{a_1} } \right ) \dots \left ( \frac{ \partial }{ i\partial s_{a_j} } + \frac{ \partial }{ \partial q_{a_j} } \right ) \left ( \frac{ \partial }{ i\partial s_{b_1} } - \frac{ \partial }{ \partial q_{b_1} } \right ) \dots \left ( \frac{ \partial }{ i\partial s_{b_k} } - \frac{ \partial }{ \partial q_{b_k} } \right ) \deth (\bs ) \nonumber \\ & & \cdot \deth (\bq ) = 8\pi^2 \left \{ (2i)^{j+2n} \frac{\partial }{ \partial r_{a_1}} \dots \frac{\partial }{ \partial r_{a_j}} \left [ \left ( \frac{\partial }{ \partial \br} \right )^2 \right ]^n \frac{ r_{c_1} \dots r_{c_{\lambda}} }{ \br^2} \left ( \frac{1 }{ \sqrt{1 - \br^2} } - 1 \right ) \right \}_{\br = 0} \nonumber \\ & & \cdot \left \{ (2i)^{k+2p} \frac{\partial }{ \partial r^*_{b_1}} \dots \frac{\partial }{ \partial r^*_{b_k}} \left [ \left ( \frac{\partial }{ \partial \br^* } \right )^2 \right ]^p \frac{ r^*_{d_1} \dots r^*_{d_{\mu}} }{ \br^{*2} } \left ( \frac{1 }{ \sqrt{1 - \br^{*2} } } - 1 \right ) \right \}_{\br^* = 0}.
\end{eqnarray}

\noindent Here
\begin{equation}                                                                  
\frac{\partial }{ i\partial s_a} + \frac{\partial }{ \partial q_a} \equiv \twi \frac{ \partial }{ \partial r_a }, ~~~
\frac{\partial }{ i\partial s_a} - \frac{\partial }{ \partial q_a} \equiv \twi \frac{ \partial }{ \partial r^*_a }
\end{equation}

\noindent and
\begin{equation}                                                                  
\frac{\partial }{ \partial r^*_a } \frac{ r_{c_1} \dots r_{c_{\lambda}} }{ \br^2} \left ( \frac{1 }{ \sqrt{1 - \br^2} } - 1 \right ) = 0, ~~~ \frac{\partial }{ \partial r_a } \frac{ r^*_{d_1} \dots r^*_{d_{\mu}} }{ \br^{*2} } \left ( \frac{1 }{ \sqrt{1 - \br^{*2} } } - 1 \right ) = 0
\end{equation}

\noindent due to analyticity (Cauchy-Riemann conditions). This is key point for the factorization to occur. Complex dummy variables $\br, \br^*$ in the RHS of (\ref{factoriz}) can equally be viewed as real independent variables, the result being the same. This looks as replacing integration over SO(3,1) by integration over SO(4). Eventually we trace back to (\ref{tilIalbegade}) where now $\bv, \br, z$ on one hand and $\bv^*, \br^*, z^*$ on another hand can be taken as independent real variables. Then $z, z^*$ on which the result $\tilIalbegade$ depends can be continued to the desired region. Therefore $2^3 \tilIalbegade (z, z^*) = \tilIalga (z) \tilIdotbedotde (z^*)$ where
\begin{eqnarray}
& & \tilIalga (z) = \int r_{c_1} \dots r_{c_{\lambda}} \left ( \frac{1 }{ \sqrt{1 - \br^2}} -1 \right ) \frac{\d^3 \br }{ 8\pi^2 \br^2} \int \exp \left (\frac{i}{ 2}\bv\br \right ) \nonumber \\ & & \cdot \frac{v^{a_1} \dots v^{a_j} }{ \rv^j} \left [ \frac{1}{ 2} \exp \frac{\rv z }{ 2i} + \frac{(-1)^j }{ 2} \exp \frac{i\rv z }{ 2} - \sum^{[j/2]}_{n=0} \frac{1 }{ (j - 2n)! } \left (\frac{\rv z }{ 2i}\right )^{j-2n}\right ] \frac{\d^3 \bv }{ \rv^2}.
\end{eqnarray}

\noindent Here integration is performed over real SO(3), $\Im \br =0, \br^2 \leq 1$, and over real $\bv$.
\vspace{-11mm} \begin{flushright} $\spadesuit$ \end{flushright} \vspace{-1mm}

Thus, the overall calculation is reduced to the separate ones in the selfdual and antiselfdual sectors and proceeds analogously to that one made in our Ref. \cite{I} in the Euclidean-like notations.

In what follows the actual case $\harcz$ is considered.

\section{Basic integrals}

According to the above said, consider only, say, selfdual (holomorphic) parts $\N_{\gamma} ( \bv )$, $\tilMalga ( l )$, $\tilIalga ( z )$ with real $\bv$, $z$. We have
\begin{eqnarray}                                                                  
\hspace{-5mm} \tilIalga ( z ) = \int \tilJal (\br , z) r_{c_1} \dots r_{c_{\lambda}} \D \pR , ~~~ \tilJal (\br , z) = \left ( \twi \right )^j \frac{\partial }{\partial r_{a_1}} \dots \frac{\partial }{\partial r_{a_j}} \tilJj (\br , z)
\end{eqnarray}

\noindent where
\begin{eqnarray}                                                                  
& & \hspace{-10mm} \D \pR = \left ( \frac{1 }{ \sqrt{1 - \br^2}} -1 \right ) \frac{\d^3 \br }{ 8\pi^2 \br^2}, ~~~ \tilJj (\br , z) = \nonumber \\ & & \hspace{-10mm} \int \exp \left (\frac{i}{ 2}\bv\br \right ) \left [ \frac{1}{ 2} \exp \frac{\rv z }{ 2i} + \frac{(-1)^j }{ 2} \exp \frac{i\rv z }{ 2} - \sum^{[j/2]}_{n=0} \frac{1 }{ (j - 2n)! } \left (\frac{\rv z }{ 2i}\right )^{j-2n}\right ] \frac{\d^3 \bv }{\rv^{j+2}}.
\end{eqnarray}

\noindent For $\tilJj$ we have equation
\begin{eqnarray}                                                                  
\left ( 2i \frac{\partial }{\partial z} \right )^j \tilJj = \int \exp \left (\frac{i}{ 2}\bv\br \right ) \left ( \cos \frac{\rv z}{2} - 1 \right ) \frac{\d^3 \bv }{\rv^2 } = - \frac{4 \pi^2}{r} \theta (z - r)
\end{eqnarray}

\noindent with initial conditions
\begin{equation}                                                                  
\tilJj^{(j - 1)} (\br , 0) = \tilJj^{(j - 2)} (\br , 0) = \dots = \tilJj (\br , 0) = 0.
\end{equation}

\noindent This gives
\begin{equation}                                                                  
(2i)^j \tilJj = - \frac{4 \pi^2}{r} \frac{(z - r)^j}{j!} \theta (z - r)
\end{equation}

\noindent and
\begin{equation}\label{I=r-dr}                                                    
\tilIalga (z ) = 4 \pi^2 \frac{(-1)^{j + 1}}{j!} \int r_{c_1} \dots r_{c_{\lambda}} \D \pR \frac{\partial }{\partial r_{a_1}} \dots \frac{\partial }{\partial r_{a_j}} \frac{(z - r)^j}{r} \theta (z - r).
\end{equation}

To study the tensor structure of $\N_{\gamma} ( \bv )$ w. r. t. the multiindex $\gamma$, we take, as above mentioned, the multiindex $\alpha$ in $\tilIalga (z )$ with $j = \lambda$. If we consider $j < \lambda$, the power of $r$ entering the integral (\ref{I=r-dr}) is such that this integral is product of $\sqrt{1 - z^2} = \cos [h / (1 + i / \gamma)]$ and of a polynomial of $z = \sin [h / (1 + i / \gamma)]$ plus, may be, $\arcsin z \propto h$. Such dependence on $h$ leads, as in lemma 1, only to the uninteresting terms in $\Ngade ( \bv , \bv^* )$ with support at $\rv^2 = 4\tilde{n}^2 (1 + i/ \gamma )^{-2}$, $n$ = 1, 2, \dots , i. e. it is equivalent to zero in the physical region. In particular, we have zero if we contract the two indices in $\gamma$, $\N_{\{ c_1 \dots c_{\lambda}\}} \Rightarrow \de^{c_{\lambda - 1} c_{\lambda }} \N_{\{ c_1 \dots c_{\lambda}\}}$ (using for studying the resulting structure the $\tilMalga ( l )$ with $\alpha$ at $j = \lambda - 2$). The requirement for the symmetrical structure composed of $\de_{ab}$ and $n_a = v_a / \rv$ to vanish when contracted over a pair of indices allows to fix this structure uniquely (up to an overall scalar factor). On the other hand, such the structure is easily written as, e. g.,
\begin{equation}                                                                  
\N_{\gamma} = N_{\lambda} \rv \frac{\partial }{\partial v^{c_1}} \dots \frac{\partial }{\partial v^{c_{\lambda}}} \frac{1}{ \rv }
\end{equation}

\noindent (at $\rv \neq 0$). To define the scalar $N_{\lambda}$, we put $\lambda = j$ and consider the moment ${\cal M}^{\alpha}_{\gamma}$ with $\alpha$, $\gamma$ contracted,
\begin{equation}\label{Malal}                                                     
{\cal M}^{\alpha}_{\alpha} (l) = C_j \int N_j (\bv^2) \, (\bv^2 )^l \d^3 \bv , ~~~ C_j \equiv \int v^{a_1 } \dots v^{a_j } \rv \frac{\partial }{\partial v^{a_1}} \dots \frac{\partial }{\partial v^{a_j}} \frac{1}{ \rv } \frac{\d^2 o_{\bsv}}{4 \pi}.
\end{equation}

\noindent By (\ref{M-I}) this is related to $\tilde{\cal I}^{\alpha}_{\alpha} (z )$. Only the lowest power of $r$ is essential in the integral (\ref{I=r-dr}) for $\tilde{\cal I}^{\alpha}_{\alpha} (z )$, the higher powers of $r$ turn out to lead, according to the above said, to the uninteresting terms in $\Ngade ( \bv , \bv^* )$ with support at $\rv^2 = 4\tilde{n}^2 (1 + i/ \gamma )^{-2}$, $n$ = 1, 2, \dots . Then we have the same constant
\begin{equation}                                                                  
\int r^{a_1 } \dots r^{a_j } r \frac{\partial }{\partial r^{a_1}} \dots \frac{\partial }{\partial r^{a_j}} \frac{1}{ r } \frac{\d^2 o_{\bsr}}{4 \pi} = C_j
\end{equation}

\noindent as in (\ref{Malal}) which factors out in $\tilde{\cal I}^{\alpha}_{\alpha} (z )$. Thus $C_j$ is canceled in the resulting equation,
\begin{eqnarray}\label{int N_j v^2l d^3 v}                                        
\int N_j (\bv^2) \, (\bv^2 )^l \d^3 \bv = \left (2i \ddh \right )^{2l+2} \frac{(-2i)^j}{(j + 1)!} \left ( \ddh \right )^{j + 1} \left [ z^{j + 1} \int\limits^z_0 \frac{-4 \pi^2}{r} \D \pR \right ] \nonumber \\ \Rightarrow \left (2i \ddh \right )^{2l+2} \frac{(-2i)^j}{(j + 1)!} \frac{\d^{j + 1} (z^{j + 1})}{\d h^{j + 1}} 2 \pi \ln \frac{1 + \sqrt{1 - z^2}}{2}.
\end{eqnarray}

\noindent Again, we have thrown away (in the last line) the terms being the derivatives of $\{ \dots z^n [\ln (1 + \sqrt{1 - z^2})]^{(k)}\}$, $n \geq k \geq 1$, the dots being a product of the derivatives of $z$. These terms result in the terms in $\Ngade ( \bv , \bv^* )$ with support at the nonphysical points $\rv^2 = 4\tilde{n}^2 (1 + i/ \gamma )^{-2}$, $n$ = 1, 2, \dots . Rescale $h \to (1 + i/\gamma)h$, then $z = \sin h$. Expand $z^{j + 1}$ over harmonics,
\begin{equation}                                                                  
\left ( \ddh \right )^{j + 1} \sin^{j + 1} h = 2^{-j} \sum^{[j/2]}_{k = 0} (-1)^k \left (^j{}^+_k{}^1 \right ) (j + 1 - 2k)^{j + 1} \cos [(j + 1 - 2k)h],
\end{equation}

\noindent $\left (^j{}^+_k{}^1 \right )$ being the binomial coefficients. Then we can use the known table integral\cite{Prud}
\begin{eqnarray}\label{sinhn ln}                                                  
\int\limits^{\infty}_0 \frac{1}{x^2 + n^2} \frac{\sh hx}{\sh \pi x} \d x = \frac{i}{2n} \sum^{n - 1}_{k = 1} \frac{(-1)^k}{n - k} e^{ihk} + i(-1)^n \frac{e^{ihn}}{2n} \ln (1 + e^{-ih}) \nonumber \\ + \frac{i}{2n!} \frac{\d^{n - 1}}{\d y^{n - 1}} \left [ \frac{(1 + y)^{n - 1}}{y} \ln (1 + y) \right ]_{y = e^{ih}}
\end{eqnarray}

\noindent to get integral representation for $\sin hn \ln \cos \frac{h}{2}$ and thus for $\cos [(j + 1 - 2k)h] \ln \cos \frac{h}{2}$. The terms other than $\sin hn \ln \cos \frac{h}{2}$ in the RHS of (\ref{sinhn ln}) lead to the usual in the present paper nonphysical terms in $\Ngade ( \bv , \bv^* )$ with support at $\rv^2 = 4\tilde{n}^2 (1 + i/ \gamma )^{-2}$, $n$ = 1, 2, \dots . The integral representation allows to represent the RHS of the Eq. (\ref{int N_j v^2l d^3 v}) in the form of its LHS, $\int N_j (\bv^2)^l \d^3 \bv$, and find
\begin{eqnarray}                                                                  
N_j = \frac{-i}{2} \left ( \frac{1}{\gamma} \! - \! i\right )^{-j} \!\!\! \frac{ \left (\frac{1}{\gamma} - i \right ) \frac{\rv}{2}}{\sh \left [\pi \! \left (\frac{1}{\gamma} \! - \! i\right ) \frac{\rv}{2} \right ]} \sum^{[j/2]}_{k=0} \frac{(-1)^k}{(j \! + \! 1)!} \frac{ (j + 1 - 2k)^{j + 1}\left (^j{}^+_k{}^1 \right )}{\left (\frac{1}{\gamma} \! - \! i\right )^2 \! \frac{\rv^2}{4} \! + \! (j \! + \! 1 \! - \! 2k)^2}.
\end{eqnarray}

\noindent  It is useful to perform explicit summation of the fractions under the sum sign reducing these to a common denominator. A number of the leading at small $\rv^2$ power terms in the resulting nominator vanish according to the equality
\begin{eqnarray}                                                                  
\sum^{[j/2]}_{k = 0} (-1)^k (j + 1 - 2k)^m \left (^j{}^+_k{}^1 \right ) = 0, ~~~ j = 2, 3, 4, \dots , \nonumber \\ m = j - 1, j - 3, \dots , j ({\rm mod} 2) + 1.
\end{eqnarray}

\noindent Only the maximal power of $\rv^2$ in the nominator survives simultaneously providing the behavior $\propto \rv^{-2}$ of this sum at large $\rv^2$ with the coefficient defined by
\begin{equation}                                                                  
\sum^{[j/2]}_{k = 0} (-1)^k (j + 1 - 2k)^{j + 1} \left (^j{}^+_k{}^1 \right ) = 2^j (j + 1)!.
\end{equation}

\noindent In overall, we arrive at
\begin{eqnarray}\label{int exp r r DR}                                            
\int \exp \left [ \frac{i}{2} \rv h( \bn \br ) \right ] r_{a_1} \dots r_{a_j} \D \pR = \frac{-i}{2} \left [ \frac{1}{2} \left ( \frac{1}{\gamma} - i\right )\right ]^{-j} \frac{\left (\frac{1}{\gamma} - i\right ) \frac{\rv}{2}}{\sh \left [ \pi \left (\frac{1}{\gamma} - i\right ) \frac{\rv}{2}\right ]} \nonumber \\ \cdot \left [ \prod^{[j/2]}_{k=0} \frac{1}{\left (\frac{1}{\gamma} - i\right )^2 \frac{\rv^2}{4} + (j + 1 - 2k)^2} \right ] \left [\left (\frac{1}{\gamma} - i\right )^2 \frac{\rv^2}{4}\right ]^{[j/2]} \rv \frac{\partial }{\partial v^{a_1}} \dots \frac{\partial }{\partial v^{a_j}} \frac{1}{ \rv }.
\end{eqnarray}

\noindent According to the derivation of Section \ref{isolating}, in the Minkowsky spacetime this only has the sense being paired with analogous $\int \exp \left [ \frac{i}{2} \rv^* h(\bn \br )^* \right ] \! $ $ \! (r_{b_1} \dots r_{b_k})^* \! $ $ \! \D \mR$ into the overall integral over $\D R$. Also remind that the considered equality holds in the integral sense being integrated with probe functions (vanishing on the finite set of nonphysical points), since the LHS is not defined as ordinary function. However, in the path integral application one needs to further integrate over bivectors, and this definition just corresponds to such possibility.

We can also introduce special definition of the Euclidean version indirectly as analytical continuation from the purely imaginary area tensors\cite{I} since the direct calculation of the path integral in the Euclidean spacetime fails because of the unboundedness of the action. We consider the Euclidean signature of space-time and SO(4) connection. Supplying (real) Euclidean variables with subscript E, the Euclidean action $S_{\rm E}$ follows formally from $S$ Eq. (\ref{S}) by substitution $\pmbvstw \to \pmbvstwE$, $\pmOmsth \to \pmOmsthE$, $\gamma \to i \gamma_{\rm E}$. Now 3-vectors $\pbvstwE$ and $\mbvstwE$ are {\it independent} variables, as well as the elements of real SO(3) $\pOmsthE$ and $\mOmsthE$ are. The appropriate path integral can be defined on the area tensor monomials by deforming integration contours to purely imaginary tensors, $\pmbvstwE \to -i\pmbvstwE$\cite{I} so that the monotonic exponent $\exp (-S_{\rm E})$ becomes oscillating one. After finding $\N$ or, more exactly, $\N_{\rm E}$ in this region, we should return to the original region of tensor values, $\pmbvstwE \to i\pmbvstwE$. Defined in this way Euclidean counterpart of the RHS of Eq. (\ref{int exp r r DR}) would read
\begin{eqnarray}                                                                  
& & \frac{1}{2} \left [ \frac{1}{2} \left ( \frac{1}{\gammaE} + 1\right )\right ]^{-j} \frac{\left (\frac{1}{\gammaE} + 1\right ) \frac{\prvE}{2}}{\sh \left [ \pi \left (\frac{1}{\gammaE} + 1\right ) \frac{\prvE}{2}\right ]} \left [ \prod^{[j/2]}_{k=0} \frac{1}{\left (\frac{1}{\gammaE} + 1\right )^2 \frac{\prvE^2}{4} + (j + 1 - 2k)^2} \right ] \nonumber \\ & & \cdot \left [\left (\frac{1}{\gammaE} + 1\right )^2 \frac{\prvE^2}{4}\right ]^{[j/2]} \prvE \frac{\partial }{\partial \pv^{a_1}_{\rm E}} \dots \frac{\partial }{\partial \pv^{a_j}_{\rm E}} \frac{1}{ \prvE }.
\end{eqnarray}

\noindent (The LHS would be $\int \exp \left [ - \frac{1}{2} \prvE h( \pbnE \pbrE ) \right ] \pr_{{\rm E}a_1} \dots \pr_{{\rm E}a_j} \D \pR_{\rm E}$. It would exponentially grow if evaluated directly.) The antiselfdual part differs by substitutions $\prvE \to \mrvE$, $\gammaE \to -\gammaE$. Here $\pmrvE = \sqrt{\pmbvE^2}$. In the physical region ($v^{ab}_{\rm E}$ is {\it bivector}) $\prvE = \mrvE \equiv {\rm v}_{\rm E}$.


\section{Conclusion}

We can mention the following properties of the basic integral (\ref{int exp r r DR}) obtained.

(i) Occurrence of the suppression factor at large area $\rvstw$
\begin{equation}\label{1/sh v}                                                    
\frac{\left (\frac{1}{\gamma} - i \right ) \frac{\rvstw}{2}}{\sh \left [\pi \left (\frac{1}{\gamma} - i\right ) \frac{\rvstw}{2} \right ]}
\end{equation}

\noindent for a triangle $\stw$ (times the same factor complex conjugated).

(ii) Regular behavior at small $\bv$. A priori we might expect for the integral of a value of the scale property of the type of $r^j$ to behave like $\rv^{-j}$. In reality, the scaling behavior of the structure $\rv \partial_{a_1} \dots \partial_{a_j} \rv^{-1} \sim \rv^{-j}$ is compensated by the factor $\rv^{2[j/2]}$, see Eq. (\ref{int exp r r DR}).

(iii) \label{item3} Good convergence properties if one calculates $\int \exp \left [ \frac{i}{2} \rv h ( \bn \br )\right ] f(\br ) \D \pR$ by expanding $f(\br )$ in the Taylor series over $\br$. Using then the result (\ref{int exp r r DR}), one gets the series for the value of interest with additional factor of the type of the inversed factorial squared, $\{[(j + 1)/2 + (i/\gamma + 1)\rv /4]![(j + 1)/2 - (i/\gamma + 1)\rv /4]!\}^{-1}$ in the order $r^j$.

The property (iii) is favorable for evaluation of the full integral over connections (\ref{N-R}) by expanding it over $\taustw$s and evaluating each term in this expansion of the type of $\int \exp \left [ \frac{i}{2} \sum_{\stw \not \in \F} \rvstw h ( \bn_{\stw} \br_{\stw} )\right ] f(\{\br_{\stw} : \stw \not \in \F \} ) \prod_{\stw \not \in \F} \D \pR_{\stw}$ (times some analogous integral over $\D \mR_{\stw}$s) by expanding $f(\{\br_{\stw} : \stw \not \in \F \} )$ in the Taylor series over $\br_{\stw}$s. Every term in these series is a product of the expressions of the type of Eq. (\ref{int exp r r DR}) over all the triangles $\stw \not \in \F$. This term includes the universal suppression factor being the product of the Eqs. (\ref{1/sh v}) over all the triangles $\stw \not \in \F$, i. e. not containing the lapse-shift edges (times the same factor complex conjugated). This is sufficient also for the suppression of the link lengths\cite{I}. Also this term has regular behavior at small $\bvstw$s.

It is an interesting question whether the result of such calculations can be expressed in some explicit closed form. Presumably it might be some special deformation of the integration contours in the Eq. (\ref{N-R}) representing it in the form of an absolutely convergent integral with the above properties.

Thus, the result of integration over connection in the Minkowsky path integral for simplicial gravity while being distribution in the configuration superspace of independent area tensors is well defined ordinary function in the physical region of bivectors. The result confirms our previous one based on evaluation of the moments of the distribution of interest in the Euclidean-like region.

The basic integrals over which the result of integration over connections in the general path integral can be expanded are evaluated. These are exponentially suppressed at large areas and regular at small ones.

\section*{Acknowledgments}

The present work was supported in part by the Russian Foundation for Basic Research
through Grants No. 08-02-00960-a and No. 09-01-00142-a.


\end{document}